\documentclass[useAMS,usenatbib]{mnras}

\usepackage{amsfonts,amsmath,amssymb,mathrsfs}
\usepackage{graphicx,epsf,epsfig}% Include figure files
\usepackage{bm}
\usepackage{longtable}
\usepackage[usenames,dvipsnames]{xcolor}
\usepackage{multirow}

\usepackage{graphicx}
\usepackage{hyperref}
\usepackage{times}
\usepackage{color}
\usepackage{breakurl}
\usepackage{color}
\usepackage{mathrsfs}

\def\be{\begin{equation}}
\def\ee{\end{equation}}
\def\bea{\begin{eqnarray}}
\def\eea{\end{eqnarray}}

\title[A model for a dark matter core at the galactic center]{A model for a dark matter core at the galactic center}

\author[K.~Boshkayev and D.~Malafarina]{K.~Boshkayev$^{1,2}$\thanks{kuantay.boshkayev@nu.edu.kz, kuantay@mail.ru} and D.~Malafarina$^2$\thanks{daniele.malafarina@nu.edu.kz}\\
$^1$NNLOT, Al-Farabi Kazakh National University, Al-Farabi av. 71, 050040 Almaty, Kazakhstan\\
$^2$Department of Physics, Nazarbayev University, Kabanbay Batyr 53, Astana 010000, Kazakhstan}

\begin{document}

\date{\today}
\maketitle

\begin{abstract}
We consider a toy model for the supermassive compact object at the galactic center that does not require the presence of a black hole. We assume a matter distribution of weakly interacting particles with a density profile inferred from dark matter profiles in the outer regions. We show that rotation curves close to the center of the Milky Way galaxy can be explained within this model. We also show that the motion of test particles (stars) at distances of the order of 100 astronomical units can not be distinguished from the motion of corresponding particles in the Schwarzschild geometry. However, differences arise at shorter distances, suggesting that it could be possible to observationally test the validity of the model in the near future.
\end{abstract}

%\pacs{04.20, 95.30, 95.30.S, 97.20.R,}


\section{Introduction}\label{sec:1}
Our modern description of the structure of galaxies relies on two important features that are not, at present, entirely understood: these are the presence of super-massive black hole candidates (SMBH) at the center of almost every galaxy and the existence of dark matter (DM) embedding and surrounding every galaxy.

Super-massive black hole candidates are generally required to explain the powerful x-ray emission observed in quasars and the nature of Active Galactic Nuclei (AGN). The existence of such objects is also validated by the observation that one super-massive compact object, called Sagittarius-A$^*$ (Sgr-A$^*$), dwells at the center of the Milky Way Galaxy (MWG).
However, it is not presently clear how such enormous black holes (some as massive as billions of Suns) could have formed so quickly in the early universe. Some SMBHs are observed at redshift $z=7.54$ and thus must have formed in less than one billion years \citep{banados}.

Throughout the years, alternative models to black holes have been suggested in a variety of contexts. Within classical General Relativity (GR), various models for extended objects have been considered (for example gravastars
\citep{grava1,grava2} and boson stars
\citep{boson1,boson2}) as well as more exotic models requiring the existence of `naked singularities'
(see for example
\citep{JMN1,JMN2,BM,APMJ})
These models consider only the gravitational effects of alternatives to BHs and investigate their observational properties in order to determine whether a black hole can be distinguished from a so-called black hole mimicker.

In the context of astrophysics, alternative models to SMBH have been suggested by several authors. For example, already in the 90s Kundt proposed the so-called burning disk model \citep{kundt}, while more recently
Ruffini et al. proposed the so-called Ruffini-Arguelles-Rueda (RAR) model
which is based on the Fermi-Dirac statistics at finite temperatures.
The peculiarity of such models is that they do not involve any SMBH at the center of any
galaxy.
For example, in the RAR model, by selecting the values of a series of physical parameters
one can reproduce the rotation curves (RCs) of the central and halo regions of different
galaxies, ranging from dwarf to big spirals, including the MWG.
In the RAR model a quantum core of almost constant density governed by
quantum degeneracy replaces the SMBH.
The application of the model to the MWG was considered in
Ref.~\citep{krut2018}.

The existence of dark matter was originally assumed in order to explain the RC of stars in the outer regions of galaxies. Its existence has been indirectly confirmed by multiple observations, such as for example, the motion of hot gas clouds in galaxy clusters and gravitational lensing (see \citep{freese} for a review). However, the true nature of dark matter remains a mystery to date, as no detection of candidate DM particles has been confirmed so far.   For a different perspective see recent articles by \citet{2013Galax...1..216C, luongo2018, Capozziello2018, 2018arXiv181005844C}. As of today, there is increasing evidence that DM must be explained with the existence of a new class of weakly interacting massive particles (WIMP).
In order to explain the observed rotation curves in numerous galaxies different density profiles for DM have been considered (see for example
\citep{marchesini2002,graham2006,chemin2011,nesti2013,siutsou2015}).
For the Milky Way Galaxy (MWG) the available data for rotation curves of stars at various distances from the center leads to a model with different DM profiles depending on the distance. Typically the gravitational field of the SMBH is believed to dominate up until a few parsecs from the center. Then the bulge is split into inner bulge and main bulge, and other density profiles are assumed for the outer parts of galaxy \citep{sofue2013}.

The possibility that DM, in the form of bosons, could form self-gravitating bound structures in galaxies has been recently studied in Ref.~\citep{levkov}.
In the present work we assume that DM exists and can form gravitationally bound clumps over short enough time scales. Therefore we model the DM distribution with a relativistic matter fluid. By making a specific choice for the DM density profile at the center of the galaxy, we reproduce the observed RCs for distant stars as well as the observed features of Sgr-A$^*$ without requiring the presence of a SMBH.
The density distribution of the DM profile is adopted as the exponential-sphere in accordance with Ref.~\citep{sofue2013}.
By fitting the RC data for the central part of the MWG, we obtain the values of the
parameters that determine the central density and the gravitational field in the inner regions.
We investigate the motion of test particle in the gravitational field of the proposed SMBH and compare with the results predicted by our DM profile. We show that, as expected, there are no noticeable differences for the motion of particles at distances of the order of 1000AU. Therefore the motion of stars such as S2 can not be used to distinguish between the two models. However differences appear at shorter distances (of the order of 100AU), as it can be seen from the behaviour of the effective potential. Therefore we expect that the future observations of the light emitted by the gas in the vicinity of the center will be able to constrain the validity of the model.
Such observations are expected to become available soon, as the Event Horizon Telescope
\citep{EHT}, will release imaging of the SMBH candidates in our galaxy and in the galaxy M87.

The paper is organized as follows: In section \ref{sec:2} we review the fitting procedures for the data of RCs in the galaxy, while in section \ref{sec:3} we construct an alternative toy model for the galactic center that fits the same data without the presence of a black hole. In section \ref{sec:4} we study the motion of test particles in the toy model and investigate how they could be used to distinguish our model from a black hole. Finally section \ref{sec:5} is devoted to a brief discussion of possible future observations that could help constrain the validity of the proposed scenario.

%%%%%%%%%%%%%%%%%%%%%%%%%%%%%%%%%%%%%%%%%%%%%%%%%%%%%%%%%%%%%%%%%%%%%%%%%%%%%%%%%%%%%%%%%%%%%%%%%%%%%%%%%%%%%%%%%%%%%%%%%%%%%%%%%%%%%%%%%%%%%%%%%%
%%%%%%%%%%%%%%%%%%%%%%%%%%%%%%%%%%%%%%%%%%%%%%%%%%%%%%%%%%%%%%%%%%%%%%%%%%%%%%%%%%%%%%%%%%%%%%%%%%%%%%%%%%%%%%%%%%%%%%%%%%%%%%%%%%%%%%%%%%%%%%%%%%
\section{Rotation curves for the Milky Way Galaxy}\label{sec:2}
\begin{figure}
\centering
\includegraphics[width=1\columnwidth,clip]{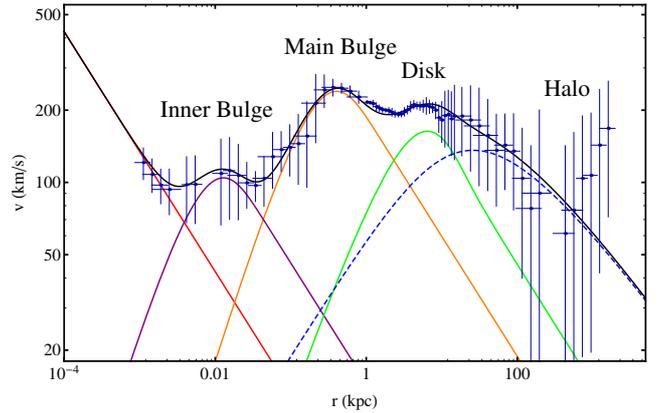}
\caption{(Color online) Rotation Curve of the Milky Way Galaxy from the data have been provided in Ref.~\citep{sofue2013}. The solid curves represent the best-fits with a central black hole, two exponential-spherical bulges, exponential thin disk, and NFW dark halo. The model parameters have been reproduced using the same fitting procedure described in Ref.~\citep{sofue2013}.}\label{fig:v_vs_r_MWG}
\end{figure}

As it is well known measuring the three dimensional velocity of a star is a challenging task even for stars within the Milky Way.
Future measurements of the proper motion of stars are expected to give a better measure of the dark matter distribution in the MWG (see \citep{gaia}).
However, as of now, there are not many references with RC data points of the MWG in the literature (see for example
\citep{clemens1985,bhattacharjee2014,huang2016}).
Also, the data vary from one source to another depending on theoretical assumptions, techniques of measurements and other factors \citep{sofue2017}.
In the following, we consider the data presented in Ref.~\citep{sofue2013}. In the article, the authors used exponential spheroidal density distributions to fit the observation data for the bulge of the MWG. %wide range of distances in the galaxy, from the bulge to the halo.
More in detail, the density profile for galactic DM presented in Ref.~\citep{sofue2013} was constructed by adding four separate profiles (inner bulge, main bulge, disk and halo). In our model we rely on the same density profiles discussed above and add one more density profile to describe the DM distribution of the core of the galaxy under the assumption that no SMBH is present.
Fig.~\ref{fig:v_vs_r_MWG} shows a logarithmic plot of the
measured rotation velocities ( - observational data with error bars) combined with the grand rotation curve of the MWG
covering the dark halo region ( - the solid black curve) \citep{sofue2009, sofue2012}.

\citet{sofue2013} in order to fit the observed RC by models has assumed the following components:
\begin{itemize}
\item[-] The central black hole with mass $M_{BH}$=4.2$\times 10^{6}M_\odot$ \footnote{The value may change slightly from one reference to another (see e.g. \citep{ghez2008,genzel2010})},  shown by a solid red curve, see Fig.~\ref{fig:v_vs_r_MWG}.
\item[-] An inner bulge with the exponential sphere density profile shown by a solid purple curve.
\item[-] The main bulge with the exponential sphere density profile shown by a solid orange curve.
\item[-] An exponential flat disk shown by a solid green curve.
\item[-] A dark halo with a Navarro-Frenk-White (NFW)  profile \footnote{ Other density profiles have been proposed to model DM in the halo region (see e.g. \citep{profiles} and \citep{profiles2})} shown by a dashed blue curve.
\end{itemize}

From the above choices, the RC, as a function of distance from the center, due to SMBH, inner and main bulges, disk and dark halo components, can be calculated as
\be
v(r)^2=v_{BH}(r)^2+v_{ib}(r)^2+v_{mb}(r)^2+v_d(r)^2+v_h(r)^2,
\ee
where $v_{BH}(r), v_{ib}(r), v_{mb}(r), v_d(r)$ and $v_h(r)$ are the
circular rotation velocities at distance $r$  of the black hole, inner bulge, main bulge, flat disk and invisible halo, correspondingly. Using the least
squares fitting procedure we have reproduced and confirmed the results of
Ref.~\citep{sofue2013} with error bars.

The assumed existence of the SMBH is responsible for the motion of stars near the galactic center and it is related to the black hole's mass via
\be
v_{BH}(r)=\sqrt{\frac{GM_{BH}}{r}},
\ee
where $G$ is Newton's gravitational constant. The BH mass is inferred by fitting the data within the range $(10^{-3}\leq r \leq 2)$ pc. At larger distances, to fit the observed RC in the bulge of the MWG the exponential sphere model has
been used. In this case, the volume mass density $\rho$ is represented by an exponential function of radius $r$ as
\be\label{eq:denprof}
\rho(r)=\rho_0 e^{-r/r_0},
\ee
where the central density $\rho_0$ and the scale radius $r_0$ are the free parameters to be fixed from observations.
This choice of the density profile, is a simplified version of the power-law with exponential cut-off profile used in Ref.~\citep{McMillan} to describe the DM density in the bulge, which is given by
\be
\rho(r)=\rho_0\left(\frac{r_1}{r}\right)^\alpha e^{-(r/r_0)^\beta},
\ee
where $\alpha$, $\beta$, $r_1$ are again model dependent parameters.

The mass enclosed within radius $r$ is given by
\be \label{eq:mass}
M(r)=M_0F(x),
\ee
where $x=r/r_0$ is the dimensionless radial coordinate,  $M_0=8\pi r_0^3
\rho_0$ is the total mass and
\be
F(x)=1-e^{-x}(1+x+x^2/2).
\ee
The circular rotation velocity in the bulge then is given by
\be
v_b(r)=\sqrt{\frac{GM(r)}{r}}=\sqrt{\frac{GM_0}{r}F\left(\frac{r}{r_0}\right)}.
\ee
For the inner and main bulges the same expression is employed with
corresponding parameters inferred from the fit of the observational data within the range $ (2 \leq r \leq 10^3)$ pc.

The expression for the flat disk and dark halo rotation velocities $v_d(r)$ and $v_h(r)$ are given in detail in Ref.~\citep{sofue2012}. The parameters of the extended disk have been calculated by fitting the RC data within the range ($1 \leq r \leq 10$) kpc. In an analogous way the dark halo parameters were found in the range ($10 \leq r \leq 400$) kpc, employing the NFW profile. It should be stressed that due to the huge error bars  of the data (related to the technical difficulties of measurements), particularly in the halo region, the fit in Fig.~\ref{fig:v_vs_r_MWG} diverges at large $r$. Throughout the paper the radial coordinate $r$ will be used in different units such as AU, pc and kpc whenever necessary to have best representations of the results. Since we are considering the possibility of having a DM distribution at the core of the galaxy in place of the SMBH the impact of the bulges, flat disk and dark halo on the motion of the test particles near the core is negligible and will not be considered here. Therefore, in the following we shall focus only on the central part of the MWG.

%%%%%%%%%%%%%%%%%%%%%%%%%%%%%%%%%%%%%%%%%%%%%%%%%%%%%%%%%%%%%%%%%%%%%%%%%%%%%%%%%%%%%%%%%%%%%%%%%%%%%%%%%%%%%%%%%%%%%%%%%%%%%%%%%%%%%%%%%%%%%%%%%%
\section{A model for a Dark Matter clump at the galactic center}\label{sec:3}

Measurements for the mass of the SMBH candidate in Sgr-A$^*$, have been used to impose upper and lower bounds to the size of the event horizon. Currently the accepted value for the mass of the SMBH candidate is $M_{BH}\approx4.2\times 10^{6}M_\odot$ (where $M_\odot$ is the mass of the Sun) while the minimum radius of the object is  $r_{BH}=2GM_{BH}/c^2\approx 0.083$ and the minimum apparent size (diameter) has been inferred to be around $\sim 5.2r_{BH}\approx 0.43$ astronomical units (AU) \citep{doeleman2008}, where $c$ is the speed of light in vacuum. The apparent diameter has been calculated from the data showing that the radio emissions from Sgr-A$^*$ are not centred on the central BH but arise from a bright spot in the region around the BH, close to the event horizon, presumably in a compact portion of an accretion disk or a relativistic jet of material that is Doppler-enhanced by its velocity along our line of sight \citep{doeleman2008}. This is due to the fact that emissions originating from a spherical surface at a given radius from a BH is strongly lensed by gravity, and presents a magnified apparent size to observers on the Earth. Hence, distant observers will measure at  least $\sim 5.2r_{BH}$ \citep{doeleman2008}.

However, there are still a plenty of open questions related to the formation, evolution, mass gain of SMBH candidates and Sgr-A$^*$ in particular.

According to \citet{iocco2015} there is evidence indicating the
presence of DM in the inner regions of the Milky Way's bulge and it seems natural to conclude that the dark matter density will increase at smaller radii. In the model presented here the center of the galaxy is described by a gravitationally bound DM clump without the presence of any SMBH.
Assuming that DM can create condensate states at the galactic center (as suggested for example in Ref.~\citep{levkov})  we consider a relativistic model for a matter distribution that can account for the observed gravitational properties of Sgr-A$^*$.

The DM distribution is described by an exponential sphere profile i.e. Eq.~(\ref{eq:denprof}) in the same manner at the inner and main bulges.
The exponential sphere model, firstly, presents a flat central DM density profile (cored), secondly, is a plausible model on the smaller scale, and, finally, is the simplest model available in the literature to describe the DM distribution in the bulge of the MWG \citep{sofue2013}. Hence the preference was given to adopt Eq.~(\ref{eq:denprof}) as a substitute to the SMBH.

The values of the parameters are fixed by fitting the data for mass of the central object and motion of nearby stars. In particular, in order to obtain the same mass as for the SMBH candidate within a few hundred astronomical units from the center, we considered three models (in figures called 'Core$^1$', 'Core$^2$' and 'Core$^3$') obtained by defining the boundary radius within which a total mass of $\approx4.2\times 10^{6}M_\odot$ must be contained (see table Tab.~\ref{tab:1} for details).
For all three cases, the fitting function becomes equivalent to one shown in
Fig.~\ref{fig:v_vs_r_MWG}. The choices for the radii within which the core matter is contained were made to check the possibility to fit the data with different constraints, if the size of a compact object were measured (known). Indeed, in all three instances it was possible to fit the current data near the galactic centre, see Fig.~\ref{fig:v_vs_r} in the next section.
At large distances (i.e. from the inner bulge outwards) the Newtonian limit for the solution is enough to account for the motion of test particles (stars) to be compared with the measured RCs.
Close to the center (i.e. in the Core region) the relativistic equations of motion for test particles are used to model the motion of stars near the SMBH candidate.

In the relativistic regime, the line element for a spherically symmetric gravitating system can be written as
\be \label{eq:metric}
ds^2=g_{ik}dx^i dx^k=e^{\nu(r)}c^2dt^2-e^{\lambda(r)}dr^2-r^2d\Omega^2,
\ee
where co-moving spherical coordinates $\{t,r,\theta,\phi\}$ have been chosen and where $d\Omega^2$ represents the line element on the unit two-sphere. The functions $\nu$ and
$\lambda$ are the metric functions to be determined from Einstein's equations once a choice for the matter model is made. In the following we choose a simple perfect fluid energy-momentum tensor given by $T^{ik}=(\rho c^2+P)u^i u^k-P g^{ik}$, where $P$ is the fluid's pressure, $u^i$ is the four-velocity and $i,k=0, 1, 2, 3$.
Einstein's equations in the presence of matter for the above metric can then be reduced to two equilibrium equations in the form
\bea
\frac{d M(r)}{dr}=4\pi r^2 \rho(r),\qquad \qquad\qquad \qquad\qquad \qquad \quad \qquad && \label{m}\\
\frac{d P(r)}{dr}=-\frac{G\left(\rho(r)c^2+P(r)\right)\left(M(r)c^2+4\pi
r^3P(r)\right)}{c^4r\left(r-2GM(r)/c^2\right)},&&\label{tov}
\eea
where $M$ represents the amount of mass enclosed within the radius $r$ and the second equation is known as the Tolman-Oppenheimer-Volkoff (TOV) equation. Then, given the density profile $\rho(r)$, the integration of equation \eqref{tov} for a given boundary condition $\{\rho(0),P(0)\}$ provides the pressure profile $P(r)$ and solves the system completely.

The metric functions $\nu$ and $\lambda$ are obtained from the density, mass and pressure from
\bea
e^{-\lambda(r)}&=&1-\frac{2GM(r)}{rc^2}, \label{lambda}\\
\frac{d\nu(r)}{dr}&=&-\frac{2}{P(r)+\rho(r)c^2}\frac{d P(r)}{dr}. \label{nu}
\eea

In order to study the motion of test particles in the above space-time one can construct the effective potential for a particle of a given energy $E$ and angular momentum $L$.
Then the effective potential for the metric Eq.~(\ref{eq:metric}) for a particle moving in the equatorial plane ($\theta=\pi/2$) has a
following form
\bea
V_{eff}&=&\frac{U_{eff}}{mc^2}=e^{-\lambda(r)}\left(1+\frac{r^2\dot\phi^2}{c^2}\right)-{\dot t}^2e^{\nu(r)-\lambda(r)}= \nonumber \\
&=&e^{-\lambda(r)}\left(1+\frac{l^2}{r^2}\right)-\epsilon^2
e^{-\nu(r)-\lambda(r)}, \nonumber
\eea
where the dot indicates a derivative with respect to the proper time and $U_{eff}$ is the total effective potential for a test particle of mass $m$. Then
$l=L/(mc)$ and $\epsilon=E/(mc^2)$ are the specific orbital momentum and the specific energy respectively \citep{ohanian2013}.
In the case of a black hole solution given by the Schwarzschild line element we have
\be
e^{\nu(r)}=e^{-\lambda(r)}=1-\frac{2GM_{BH}}{rc^2},
\ee
and the specific effective potential becomes
\be
V_{eff}=\left(1-\frac{2GM_{BH}}{rc^2}\right)\left(1+\frac{l^2}{r^2}\right)-\epsilon^2.
\ee
%
%--{\it Newtonian case}.
In the limiting case $c\rightarrow\infty$ we retrieve the Newtonian limit and the
expressions for the mass, pressure and gravitational potential become
\bea \label{eq:classequil}
\frac{d M(r)}{dr}&=&4\pi r^2 \rho(r),\\
\frac{d P(r)}{dr}&=&-\rho(r)\frac{GM(r)}{r^2},\\
\frac{d\Phi(r)}{dr}&=&-\frac{1}{\rho(r)}\frac{d P(r)}{dr}.  \eea
where $\Phi(r)$ is the gravitational potential within the core, which is
related to the dimensionless potential (metric function) via
$\nu(r)=2\Phi(r)/c^2$.

For numerical computations it is convenient to use dimensionless forms of the
above equations by taking the parameter $r_0$ in the density profile
\eqref{eq:denprof} as the unit length, therefore setting
\bea \label{eq:dimensionlessequil}
r&=&r_0 x,\\
\rho(r)&=&\frac{c^2}{G r_0^2}\rho^*(x),\\
m(r)&=&\frac{c^2 r_0}{G}m^*(x),\\
P(r)&=&\frac{c^4}{G r_0^2}P^*(x),\\
\Phi(r)&=&c^2\Phi^*(x),
\eea
where the asterisk indicates corresponding dimensionless quantities.
%In case of necessity one can always recover physical quantities.
%
%%%%%%%%%%%%%%%%%%%%%%%%%%%%%%%%%%%%%%%%%%%%%%%%%%%%%%%%%%%%%%%%%%%%%%%%%%%%%%%%%%%%%%%%%%%%%%%%%%%%%%%%%%%%%%%%%%%%%%%%%%%%%%%%%%%%%%%%%%%%%%%%%%

\section{Distinguishing the Dark Matter core from a Black Hole}\label{sec:4}

\begin{figure}
\centering
\includegraphics[width=1\columnwidth,clip]{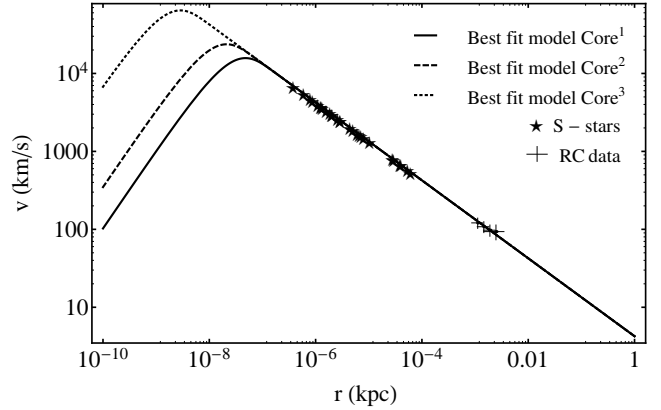}
\caption{Rotation Curve of the central part of the Milky Way Galaxy. The solid curve corresponds to the best fit with no constraints on the mass (Core$^1$). The dashed curve shows the best fit model when 4.2$\times 10^{6}M_\odot$ mass is constrained within the radius of 120 AU (Core$^2$). The dotted curve indicates the best fit model for the same mass but constrained within 100 $r_{BH}$ $\approx8.3$ AU (Core$^3$).}\label{fig:v_vs_r}
\end{figure}

\begin{table*}
\centering
\begin{tabular}{l l l l}
\hline
Mass component &  Total mass ($M_{\odot}$) & Scale radius (pc) & Central density ($M_{\odot}/$pc$^3$)  \\
\hline
Core$^1$       & 4.2$\times10^6$                & 1.417$\times10^{-5}$        & 5.873$\times10^{19}$ \\
Core$^2$       & 4.2$\times10^6$                & 6.261$\times10^{-6}$        & 6.808$\times10^{20}$ \\
Core$^3$       & 4.2$\times10^6$                & 8.466$\times10^{-7}$        & 2.754$\times10^{23}$ \\
Inner bulge    & 5.0$\times10^7$                & 3.8                         & 3.6$\times10^4$   \\
Main bulge     & 8.4$\times10^9$                & 120                         & 1.9$\times10^2$   \\
Disk           & 4.4$\times10^{10}$             & 3$\times10^{3}$             & 15.0 \\
Dark halo      & 5.0$\times10^{10} (r \leq h)$  & $h$=12$\times10^{3}$        & 1.1$\times10^{-2}$ \\
\hline
\end{tabular}
\caption{Parameters for the exponential sphere model. The model parameters of Core$^1$ have been inferred from the fit of the RC data with no constraints on mass. For Core$^2$ the parameters have been obtained when 4.2$\times10^6 M_{\odot}$ is constrained within 120AU and for Core$^3$ the parameters have been extracted when 4.2$\times10^6 M_{\odot}$ is packed in 100$r_{BH}$. The numeric values for the inner bulge, main bulge, thin disk and dark halo have been taken from \citep{sofue2013} for clarity. Note that 1$M_{\odot}/$pc$^3 \approx6.77\times10^{-23}$g/cm$^3$, 1erg $\approx$ 624.15GeV and  1pc $\approx$ 206265AU.}\label{tab:1}
\end{table*}

By fitting the RC of the MWG we inferred the model parameters $r_0$ and $\rho_0$ which are listed in Tab.~\ref{tab:1}. As said, for different constraints on the mass and size of the DM core we obtain different values for the central density of the DM clump, that still fit the RC of observed stars. In Fig.~\ref{fig:v_vs_r}, we show the RC for three different choices for the Core profile at the center of the MWG. Different constraints produce different outcomes for RCs at distances shorter than $10^{-5}$pc. Nonetheless, all of them overlap with the currently available data and can be used to explain the motion of S-stars. The data for the S-stars have been reduced from \citep{gillessen2009}.
\begin{figure*}
\centering
\begin{tabular}{lr}
\includegraphics[width=\columnwidth,clip]{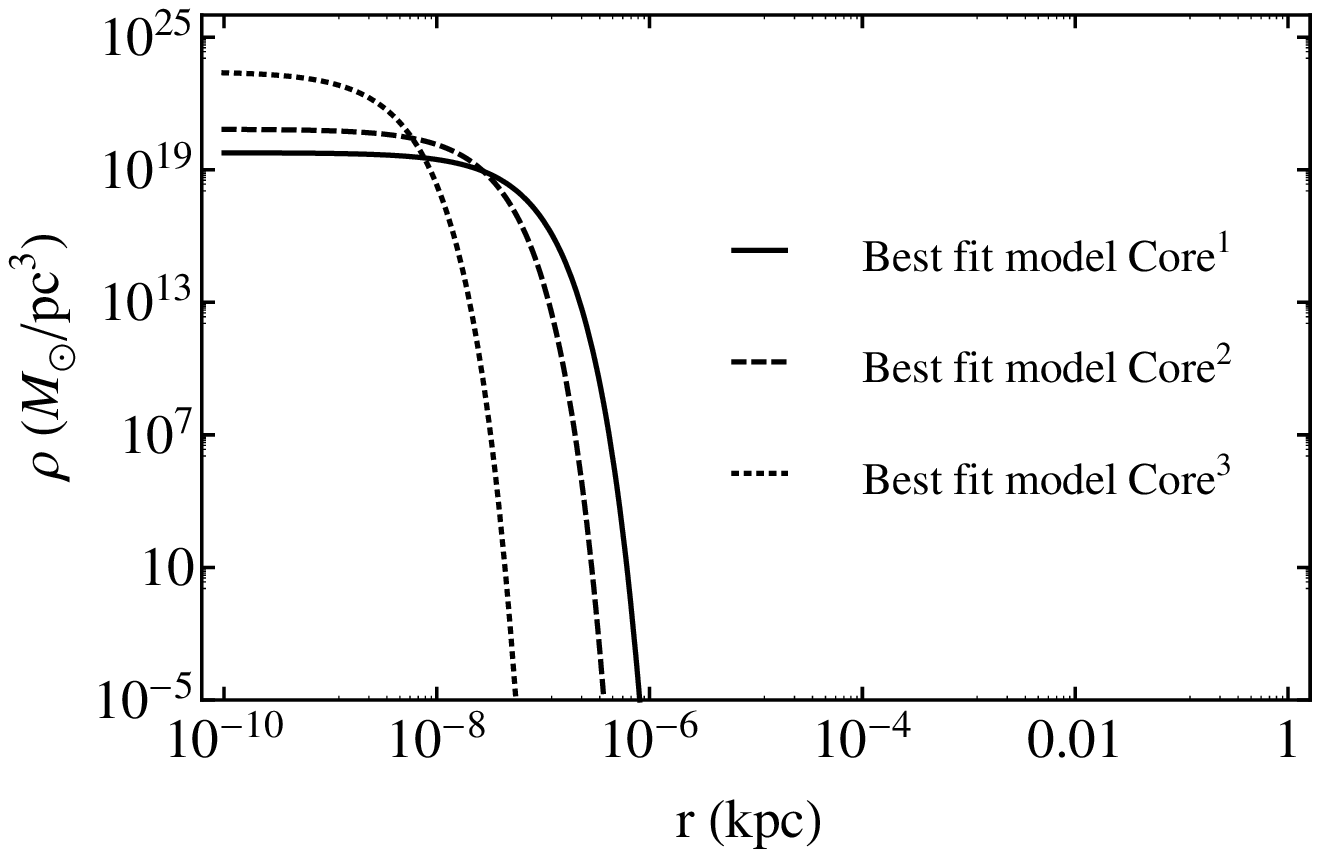} & \includegraphics[width=\columnwidth,clip]{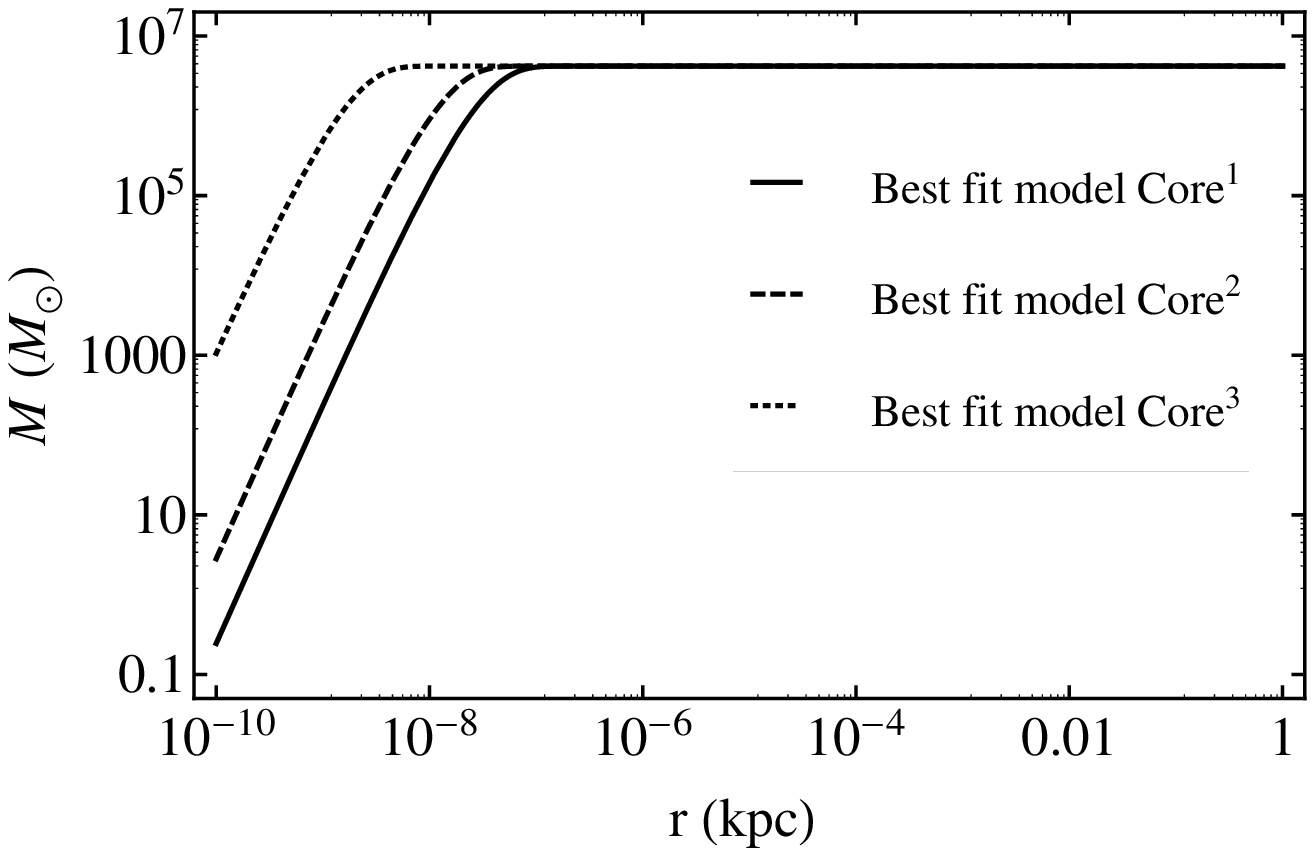}
\end{tabular}
\caption{Left panel: Density profile versus distance for the DM core. It is easy to see that the density of the DM clump becomes negligible at distances above $10^{-3}$pc, where the other DM profiles become dominant. Right panel: Mass of the DM clump as a function of distance from the center for the three different models (Core$^1$, Core$^2$ and Core$^3$). It is easy to see how the total mass flattens around the measured mass for the SMBH candidate within the fixed radial parameter. The legend is the same as in Fig.~\ref{fig:v_vs_r}.}\label{fig:rho_vs_r}
\end{figure*}

The Left panel of Fig.~\ref{fig:rho_vs_r} shows the density profiles of the DM core that have been used in place of the SMBH candidate. Here the profiles display distinctive features depending on the imposed constraints. The corresponding values of the central density are given in Table~\ref{tab:1} and can vary from $3.97\times10^{-3}$ g/cm$^3$ for model Core$^1$ to $18.64$ g/cm$^3$ for model Core$^3$.
The Right panel of Fig.~\ref{fig:rho_vs_r} shows the mass profile for the core of the MWG. All constraints and henceforth models Core$^1$, Core$^2$ and Core$^3$ yield the same total mass that is equal to the SMBH mass $M_{BH}\approx M_0$ (see Tab.~\ref{tab:1}).

\begin{figure*}
\centering
\begin{tabular}{lr}
\includegraphics[width=\columnwidth,clip]{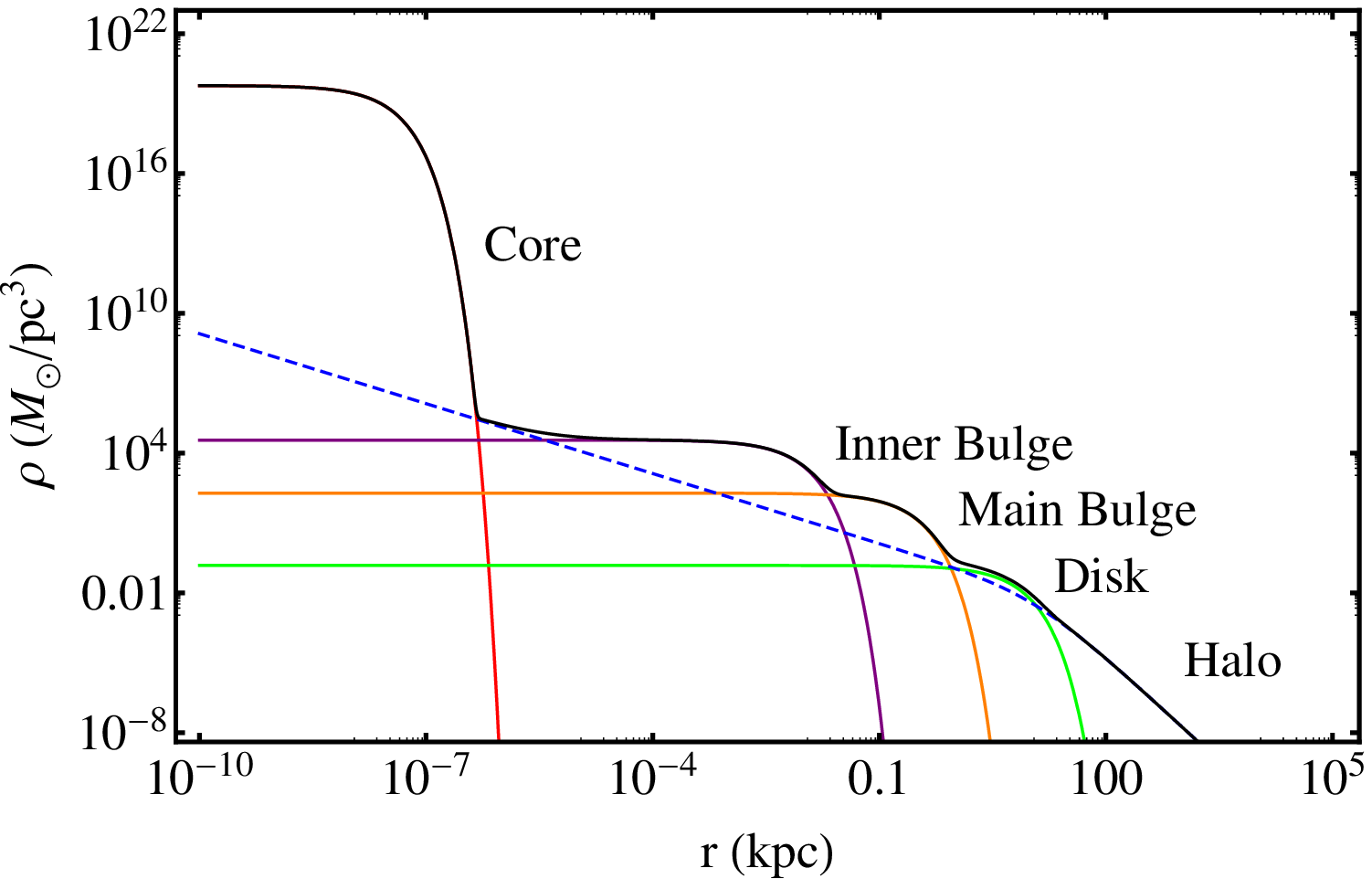} & \includegraphics[width=\columnwidth,clip]{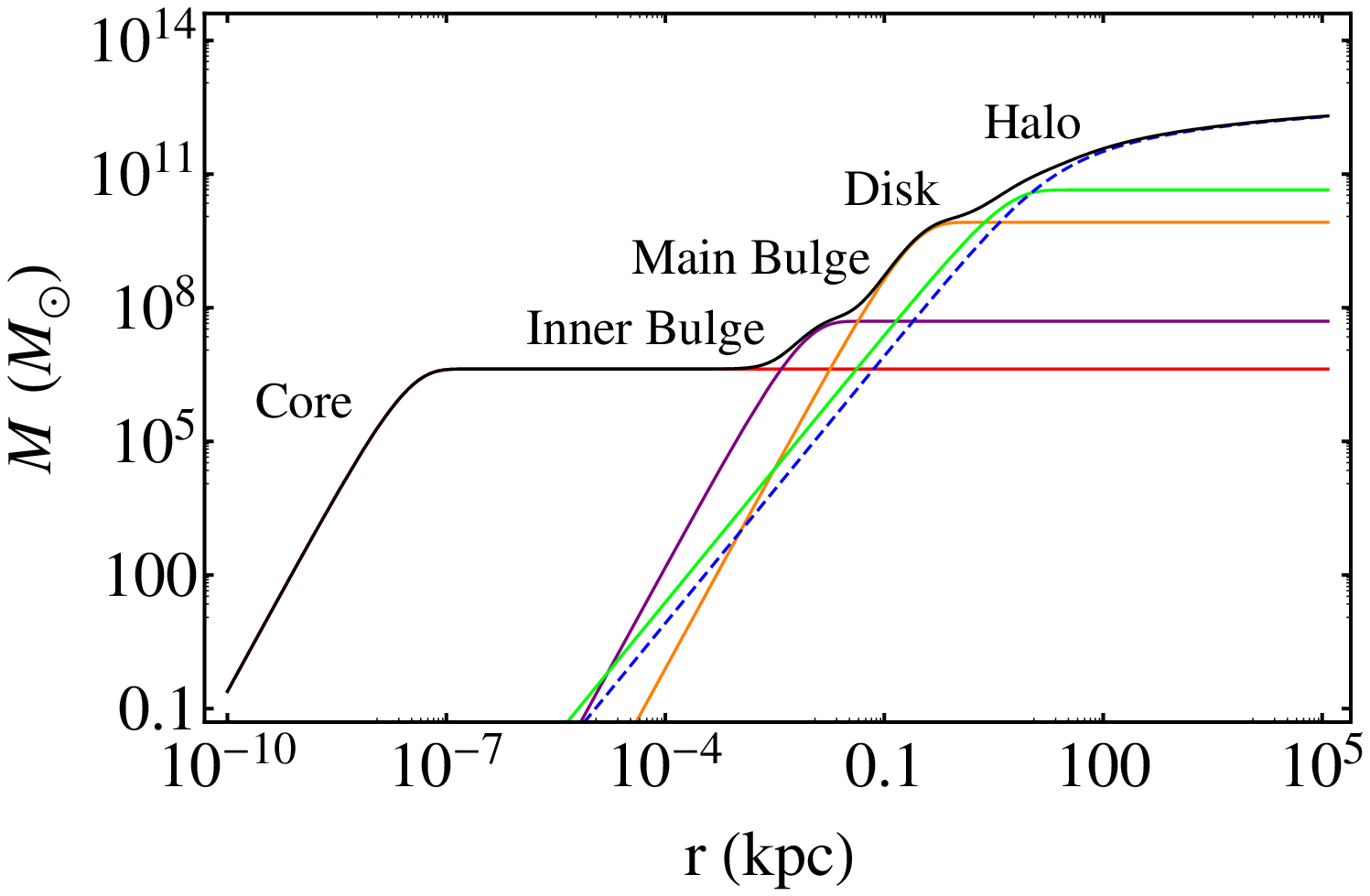}
\end{tabular}
\caption{(Color online) Left panel: Logarithmic plots of the volume density of the core, exponential spheroids, exponential disk, and NFW halo calculated for the fitted parameters. The Core$^1$ model has been employed hereafter in the paper for the sake of generality. Right panel: Mass distribution within the MWG for the proposed model. The values for the total mass of each constituent are given in Tab.~\ref{tab:1}.}\label{fig:rho_vs_r_MWG}
\end{figure*}

In the following we shall restrict the analysis of orbits for the model given by the scale radius of Core$^1$. However, similar conclusions can be derived for other models with different central densities as constrained by possible future observations.
In the Left part of  Fig.~\ref{fig:rho_vs_r_MWG} the density profile of Core$^1$ model has been used to model the galactic core instead of the SMBH candidate. The DM profiles for the outer regions overlap with the ones described in Ref.~\citep{sofue2013}.

In the Right part of Fig.~\ref{fig:rho_vs_r_MWG} the mass profile for the MWG is shown with each individual constituent from the core to the dark halo. It is interesting to notice that, as expected, the main contribution to the mass of the MWG is made by the distribution of DM in flat, thin disk and dark halo and the presence of a SMBH or a DM clump bear no effect at distances larger than  $1$pc.

To investigate the motion of test particles in the gravitational field of massive source described by our density profile one needs to solve the mass balance equation, Eq.~\eqref{m}, together with the TOV equation, Eq.~\eqref{tov}, including the equation for the dimensionless gravitational potential, Eq.~\eqref{nu}. The mass distribution is found by integration Eq.~\eqref{m}, fulfilling the regularity condition $M(0)=0$ at the origin, which yields Eq.~(\ref{eq:mass}) in an analytic form in both relativistic and classical cases.

\begin{figure*}
\centering
\begin{tabular}{lr}
\includegraphics[width=\columnwidth,clip]{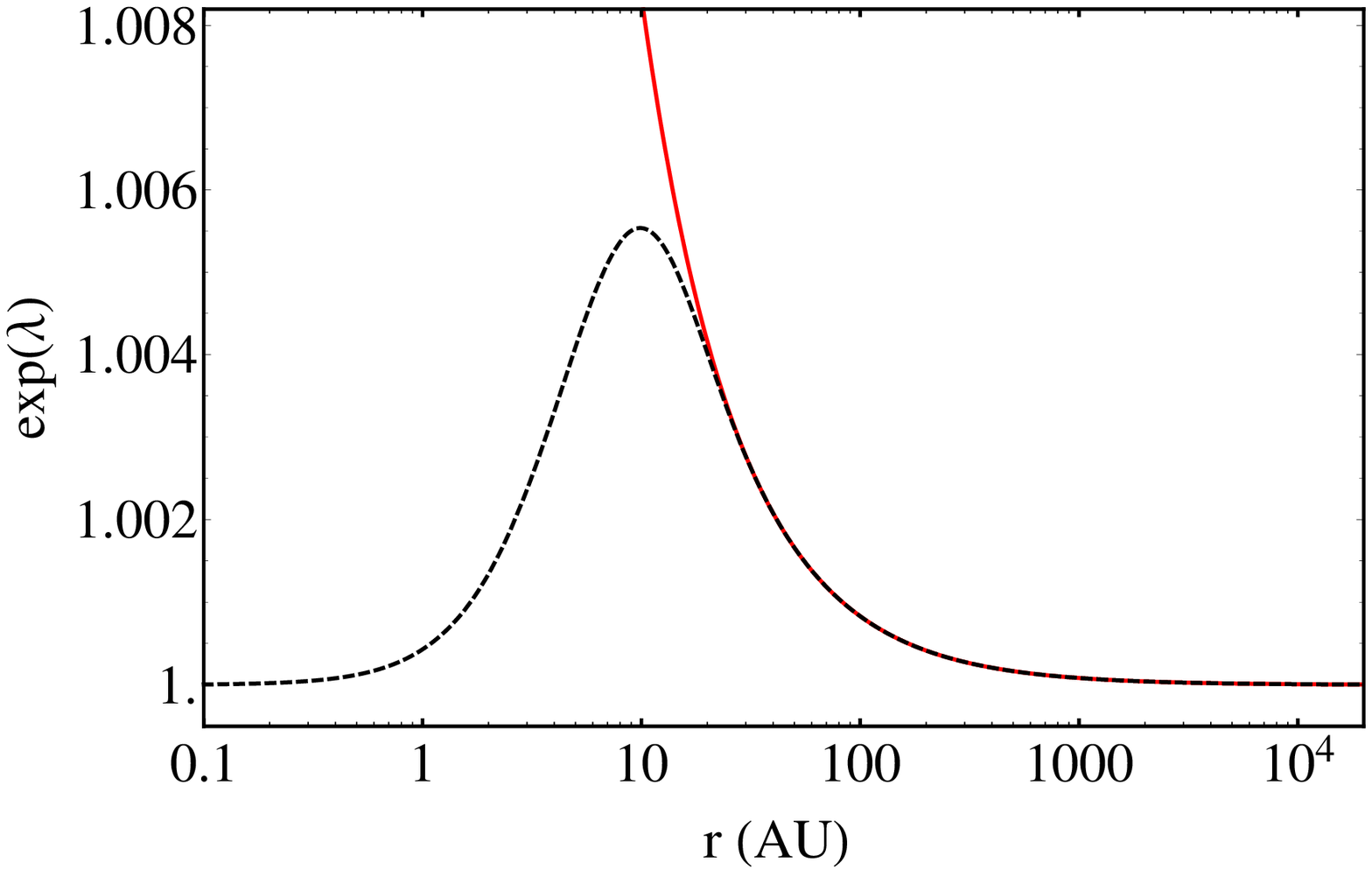} & \includegraphics[width=\columnwidth,clip]{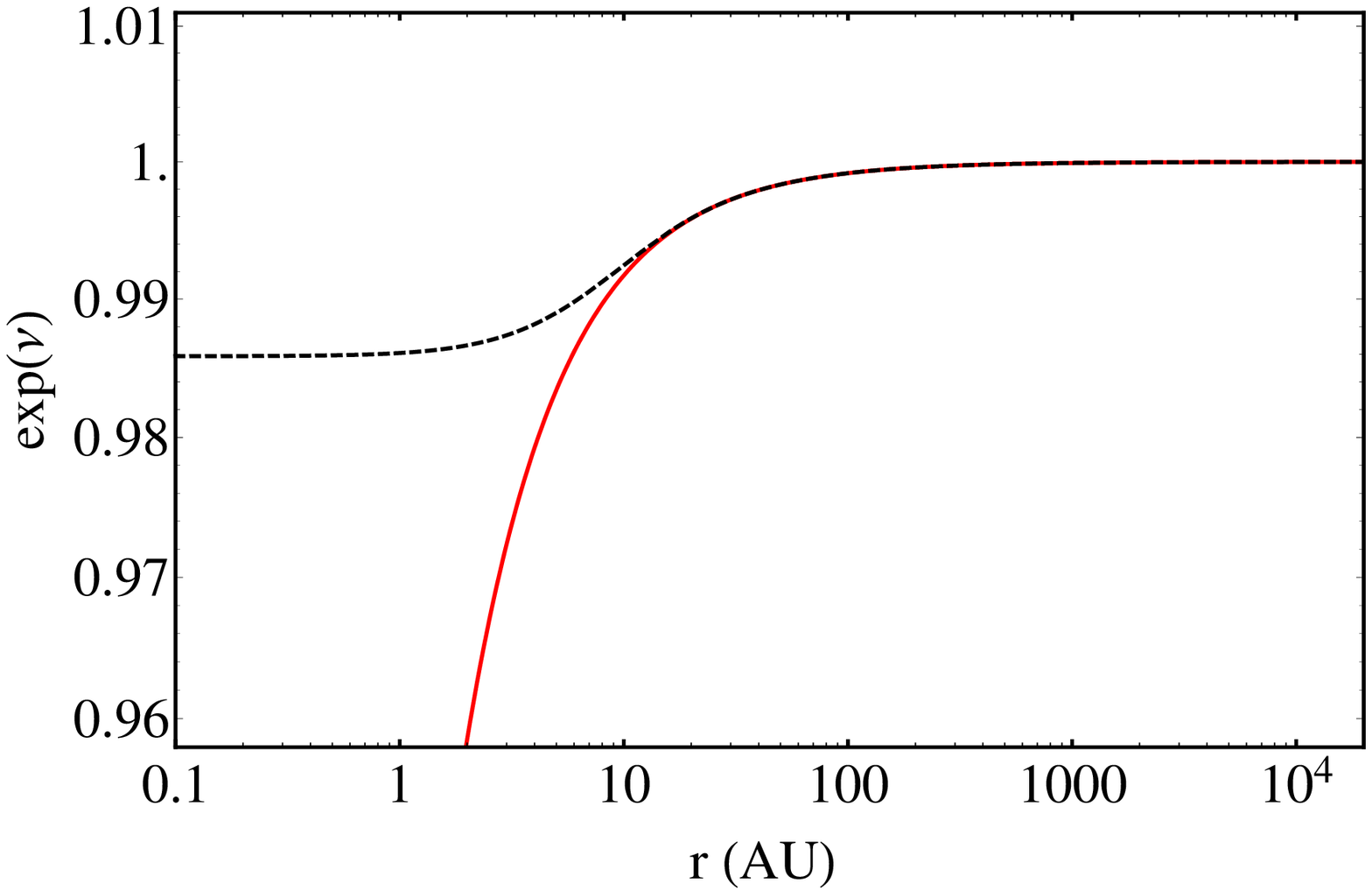}
\end{tabular}
\caption{(Color online) Left panel: The metric function $\lambda(r)$ versus radial distance $r$. Right panel: The metric function $\nu(r)$ versus radial distance $r$. In both panels, the solid red curve corresponds to the functions for the SMBH while the dashed black curve corresponds to the DM core model.}\label{fig:lambda_vs_r}
\end{figure*}

Furthermore, metric function $\lambda(r)$ is straightforwardly determined by Eq.~(\ref{eq:mass}).  The behavior of function $\lambda(r)$ near the galactic center is depicted in the Left part of Fig.~\ref{fig:lambda_vs_r} for both cases: in the presence of the SMBH and DM core. For the SMBH $\lambda(r)$ tends to infinity as one approaches the Schwarzschild radius $r\rightarrow r_{BH}$. Instead, in the presence of a DM cloud near the center $\lambda(r)$ reaches a maximum and then tends to a constant value.

After obtaining the pressure function from Eq.~\eqref{tov} one can calculate $\nu(r)$ numerically for the DM profile and compare it with its analytic counterpart for the SMBH.
The Right part of Fig.~\ref{fig:lambda_vs_r} shows metric function $\nu(r)$ versus
$r$ in AU. For the SMBH $\nu(r)$ tends to zero as it approaches the Schwarzschild radius and
for the DM core it reaches a positive constant value. For both metric functions, at radii larger than $100$AU the two functions exhibit almost identical behaviour.

\begin{figure*}
\centering
\begin{tabular}{lr}
\includegraphics[width=\columnwidth,clip]{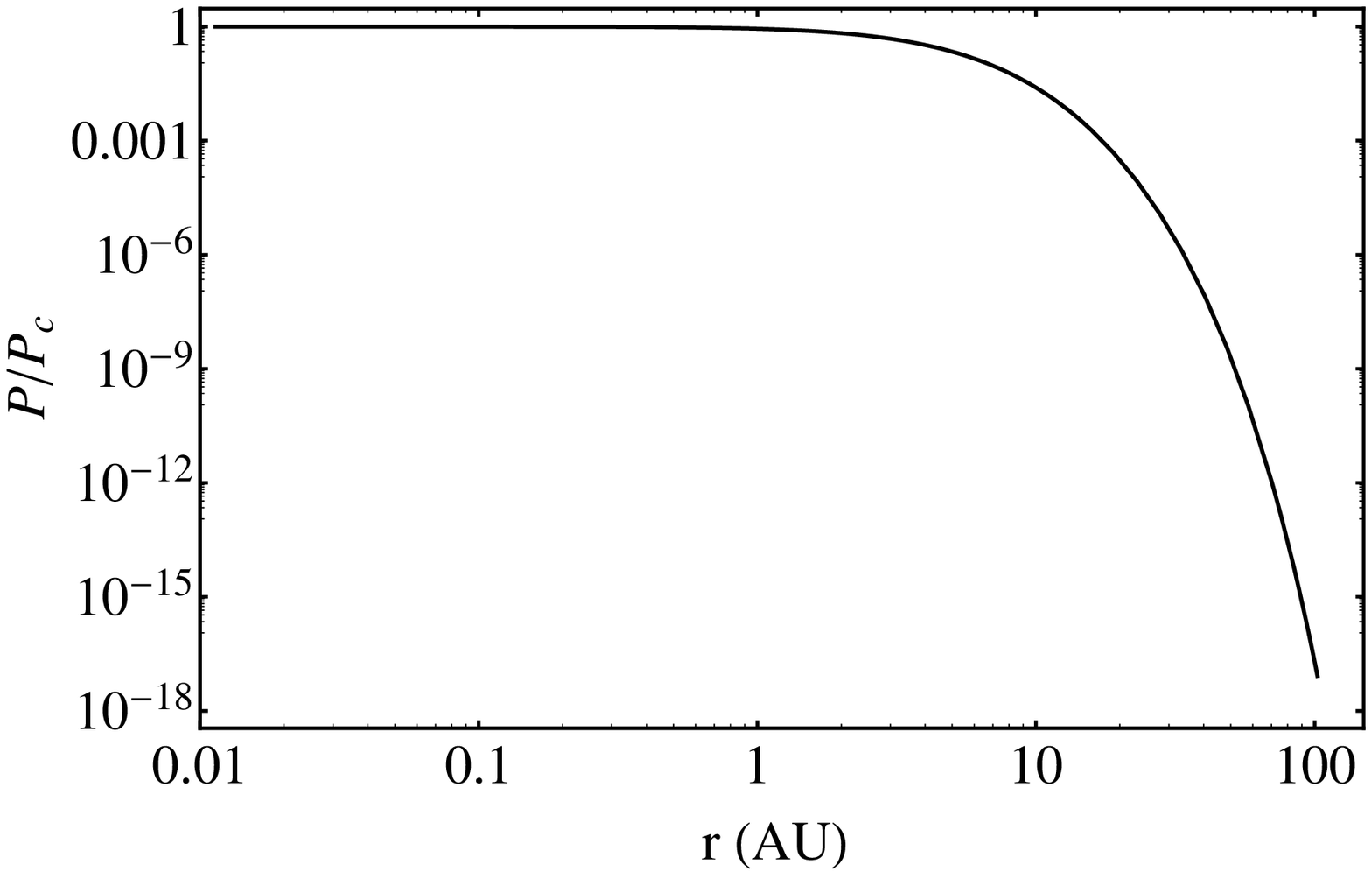} & \includegraphics[width=\columnwidth,clip]{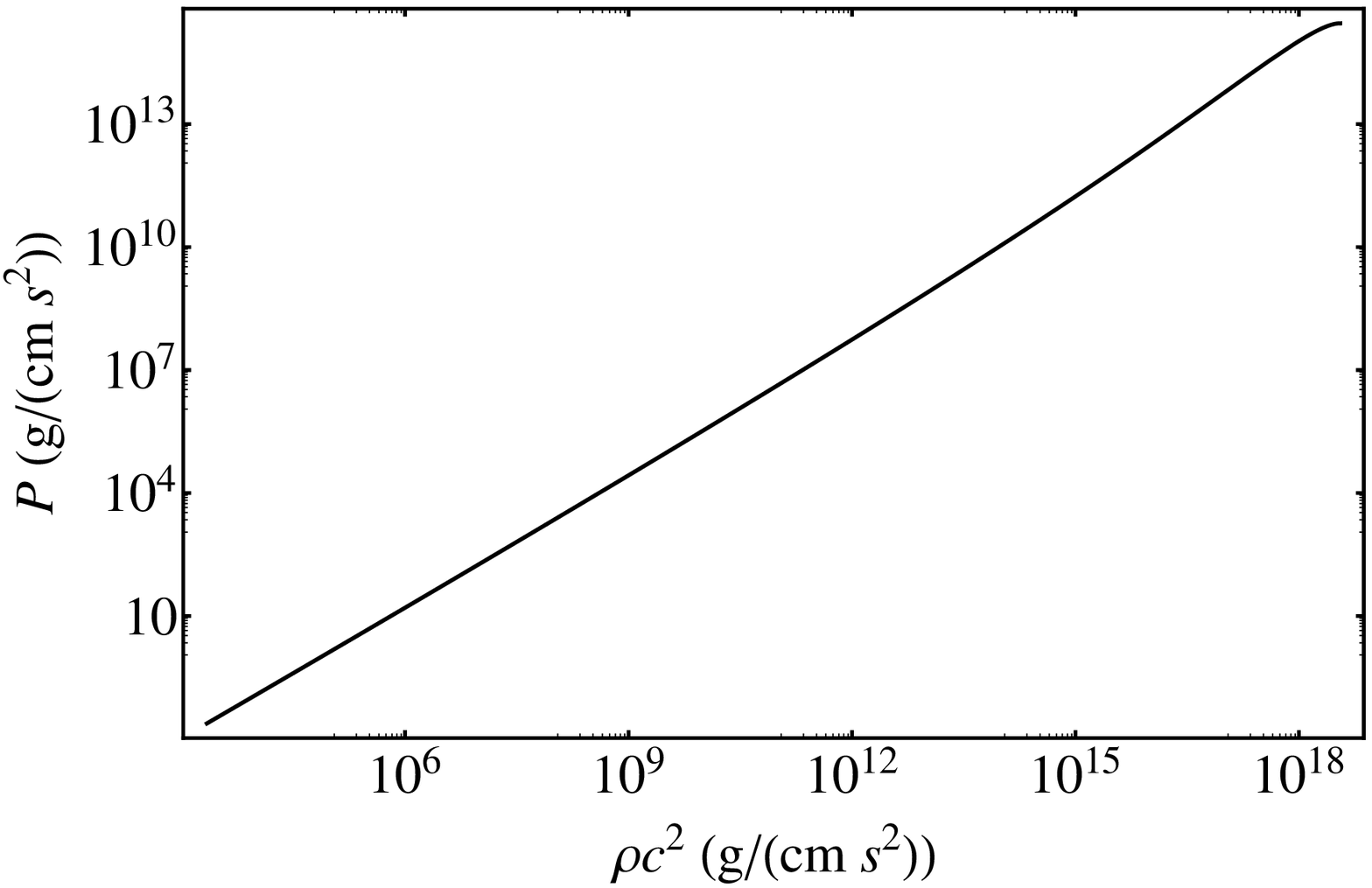}
\end{tabular}
\caption{Left panel: Pressure versus radial distance for the DM distribution as obtained from the integration of the TOV equation. Right panel: Equation of state of the dark matter core.}\label{fig:pressure_vs_r}
\end{figure*}

Of course in general the solution of
the relativistic hydrostatic equilibrium equation for pressure $P(r)$ cannot be obtained
analytically. The result of the numerical integration is illustrated in the Left panel of Fig.~\ref{fig:pressure_vs_r}, where the DM pressure profile near the MWG center is given as a function of the radial distance $r$ in AU and $P_c=2.88\times10^{15}$ g$/$(cm s$^2$) is the value of the central pressure.

To construct the plots for the DM pressure and metric function $\nu$ the hypothetical surface of the galactic core was defined at the radius where the core density profile intersects the inner bulge density profile (see the left panel of Fig.~\ref{fig:rho_vs_r_MWG}). This procedure is a necessary step to exclude the impact of outer parts of the Galaxy (see Tab.~\ref{tab:1}). The corresponding radius of the core is then equal to 102.4 AU.  At that radius the density and pressure drop to $\sim 10^{-16}$ and $\sim 10^{-18}$ times their central values, respectively.

It is important to stress that in the Newtonian case, unlike in GR, the solution of the system of equilibrium equations for the density profile in Eq.~(\ref{eq:denprof})
is analytic. The expressions for the pressure and
gravitational potential can be obtained by satisfying the conditions at
infinity and they can be written as
\bea
P(r)&=&8\pi G r_0^2 \rho_0^2K(x), \label{eq:pressN}\\
\Phi(r)&=&-8\pi G r_0^2 \rho_0 N(x).
\eea
where
\bea
K(x)&=&\frac{e^{-x}}{x}-e^{-2x}\left(\frac{1}{4}+\frac{1}{x}\right)+{\rm Ei}(-x)-{\rm Ei}(-2x), \nonumber\\
N(x)&=&\frac{1}{x}-e^{-x}\left(\frac{1}{2}+\frac{1}{x}\right), \nonumber
\eea
and ${\rm Ei}(x)$ is the exponential integral for real non zero values of $x=r/r_0$
defined as
\be
{\rm Ei}(x)=-\int_{-x}^\infty\frac{e^{-t}}{t}dt.
\ee
The values of the pressure and potential at the center in the Newtonian case are
\bea
P_c&=&P(0)=2\pi G r_0^2 \rho_0^2(3-\ln16),\\ \nonumber
\Phi_c&=&\Phi(0)=-4\pi G r_0^2 \rho_0.\nonumber
\eea
These values are crucial as they serve to check the correctness of the numerical results in the relativistic case. Indeed, the difference between the relativistic numerical and classical analytic results for the pressure and potential is minimal. Hence with a higher accuracy one can safely study the DM in the Newtonian framework.

It is worth noticing that Eq.~(\ref{eq:pressN}) together with Eq.~(\ref{eq:denprof}) provide an equation of state for the DM core of the MWG, as can be seen in the right panel of Fig.~\ref{fig:pressure_vs_r}. The relation between the pressure and density is almost linear and it is given as $P\approx(10^{-3}-10^{-5})\rho c^2$ from the core center to the inner bulge. Analogous result was found for DM halos of other galaxies in Ref.~\citep{barranco2015}, where the dependence was $P\approx(10^{-6}-10^{-8})\rho c^2$.

We turn now the attention to the motion of test particles in the DM cloud at the core of the galaxy. Assuming that DM particles interact mostly via gravitation with normal matter and that the densities in the core are sufficiently low to avoid collisions we can consider geodesic motion in the interior solution given by the metric in Eq.~\eqref{eq:metric}, with density profile given by Eq.~\eqref{eq:denprof} and pressure obtained from the numerical integration of Eq.~\eqref{tov}.
\begin{figure*}
\centering
\begin{tabular}{lr}
\includegraphics[width=\columnwidth,clip]{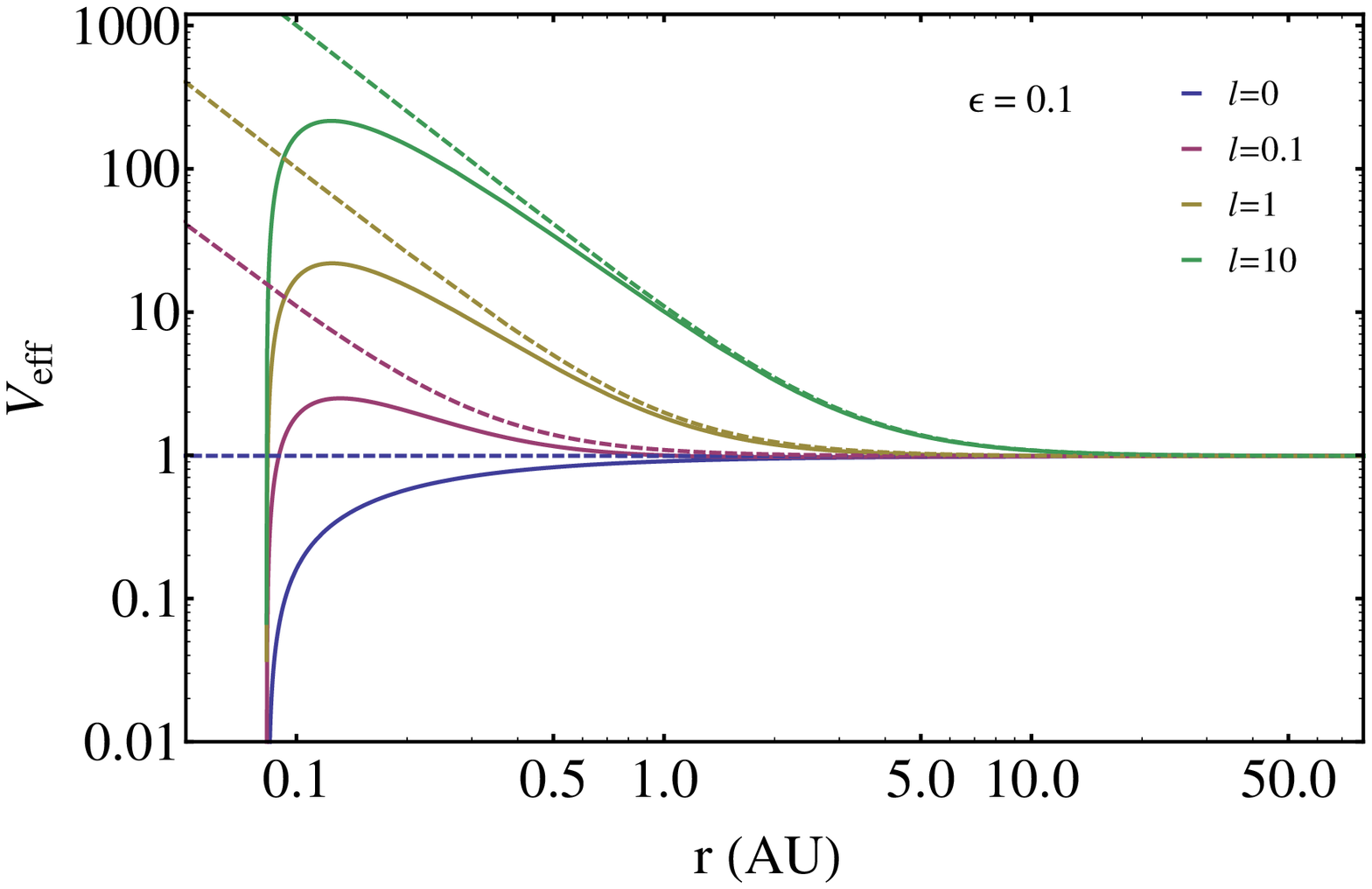} & \includegraphics[width=\columnwidth,clip]{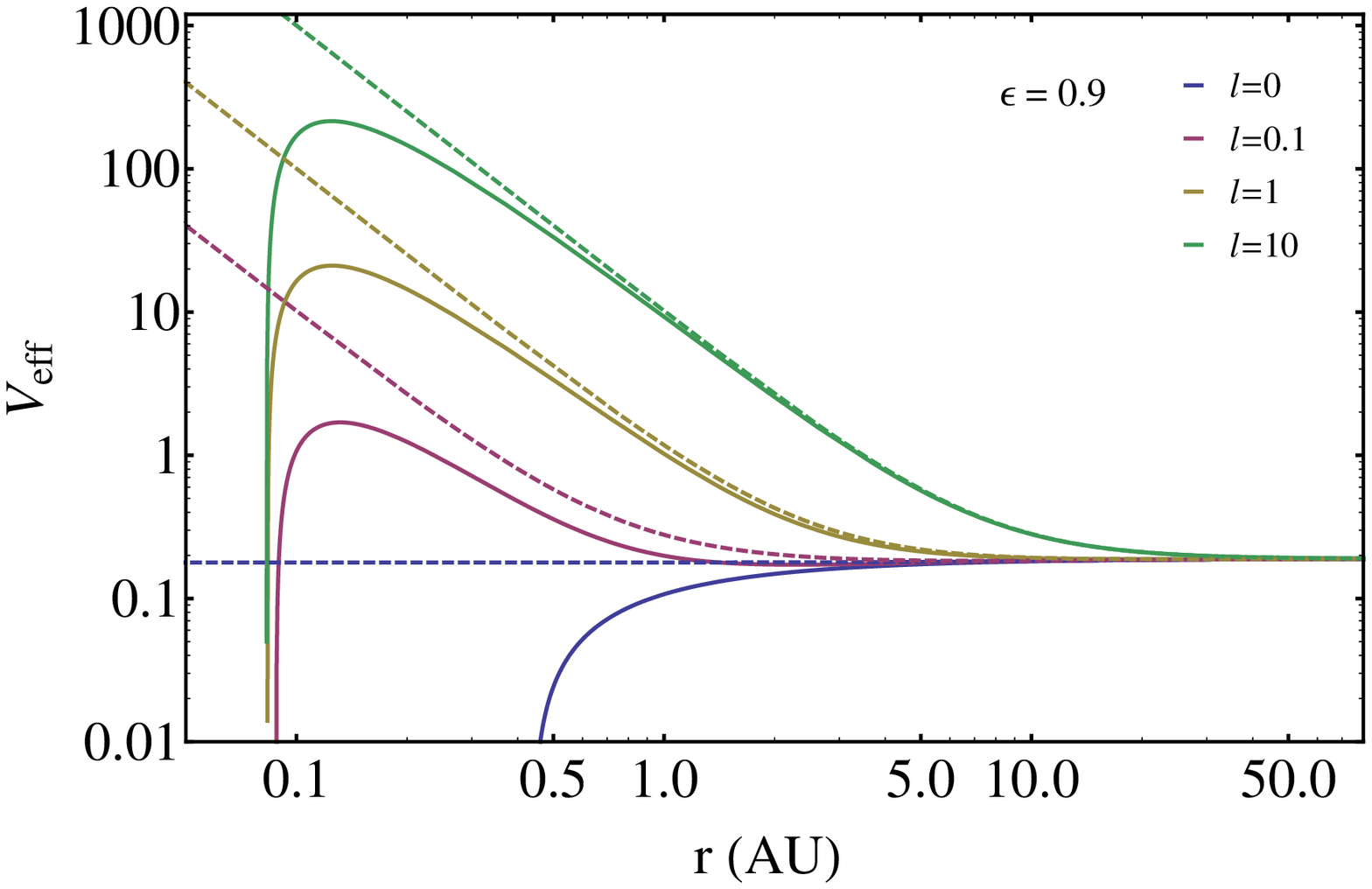}
\end{tabular}
\caption{(Color online) Specific effective potential versus distance. Solid curves correspond to $V_{\rm eff}$ of the SMBH and dashed curves to the DM, close to the center of the MWG. Left panel: The specific energy of the test particle $\epsilon$ is 0.1. Right panel: The specific energy of the test particle $\epsilon$ is 0.9.}\label{fig:Veff1}
\end{figure*}

In Fig.~\ref{fig:Veff1} it is shown the behavior of the specific effective potential of a test particle for various values of $l$ and $\epsilon$. The main difference appears close to the center where the strong field regime dominates in the presence of the SMBH. In the case of a DM cloud the gravitational field is always weak as the density remains low also for small radii.

\begin{figure*}
\centering
\begin{tabular}{lr}
\includegraphics[width=\columnwidth,clip]{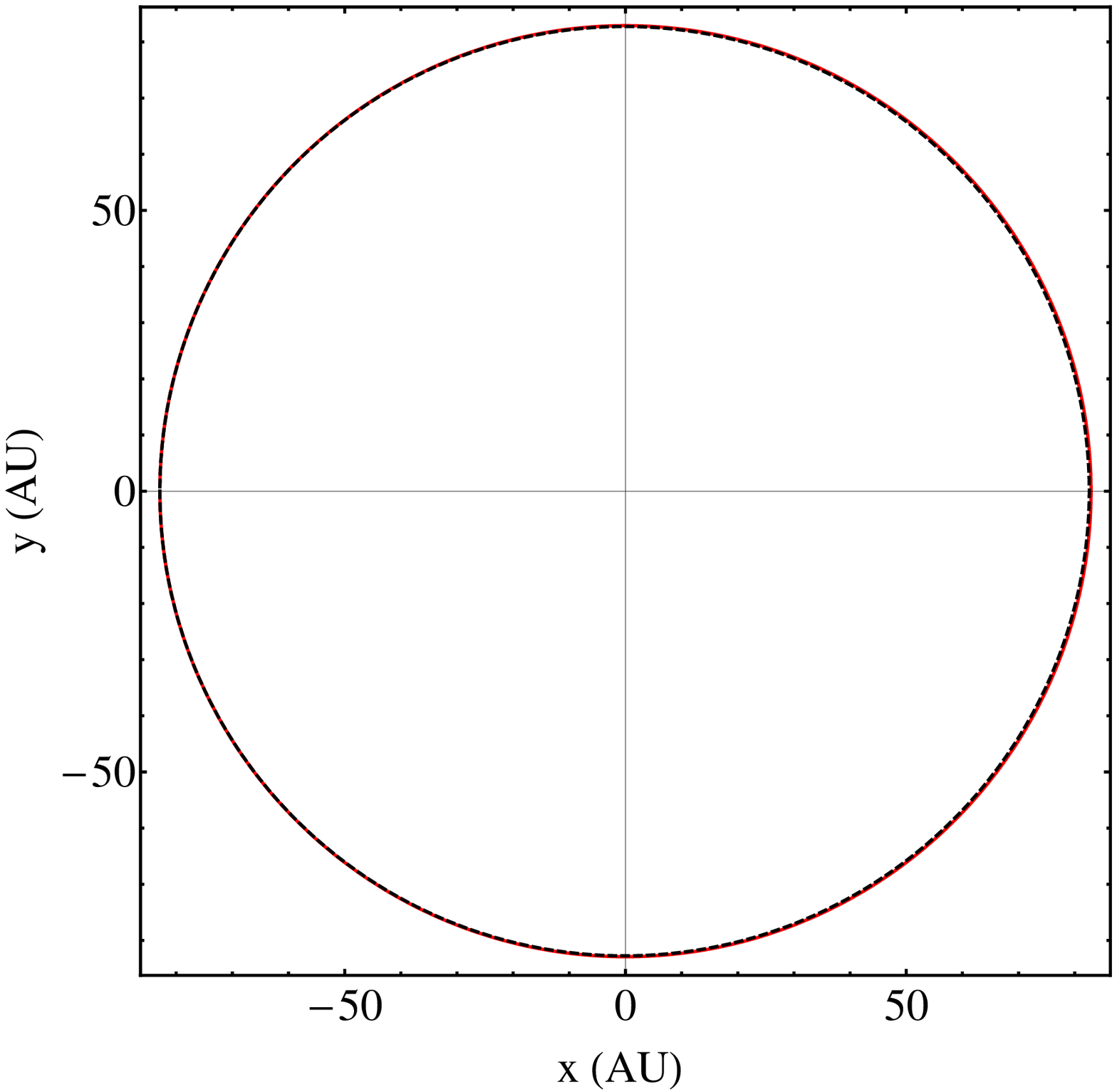} & \includegraphics[width=\columnwidth,clip]{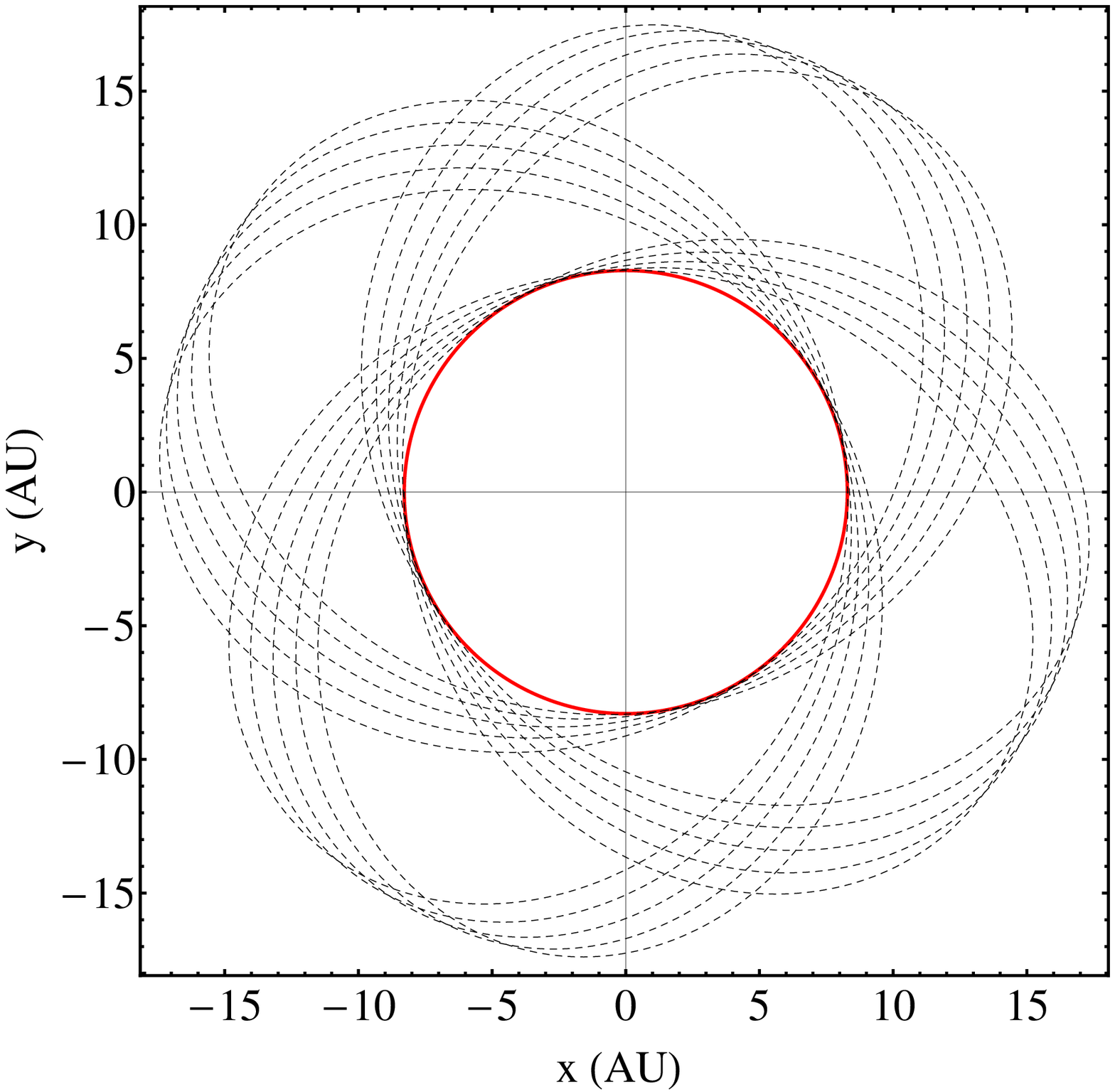}
\end{tabular}
\caption{(Color online) Geodesics near the galactic center. The solid red and the dashed black curves show the motion of test particles in the field of the SMBH and DM, respectively. Initial conditions are the same for both cases.  Left panel: $r(0)=1000r_{BH}\approx83 AU, \dot r(0)=0, \phi(0)=0, \dot\phi(0)\approx5.4\times10^{-7}$ rad$/$s. Right panel: $r(0)=100r_{BH}\approx8.3 AU, \dot r(0)=0, \phi(0)=0, \dot\phi(0)\approx1.7\times10^{-5}$ rad$/$s.}\label{fig:y_vs_x_1000RBH}
\end{figure*}

\begin{figure*}
\centering
\begin{tabular}{lr}
\includegraphics[width=\columnwidth,clip]{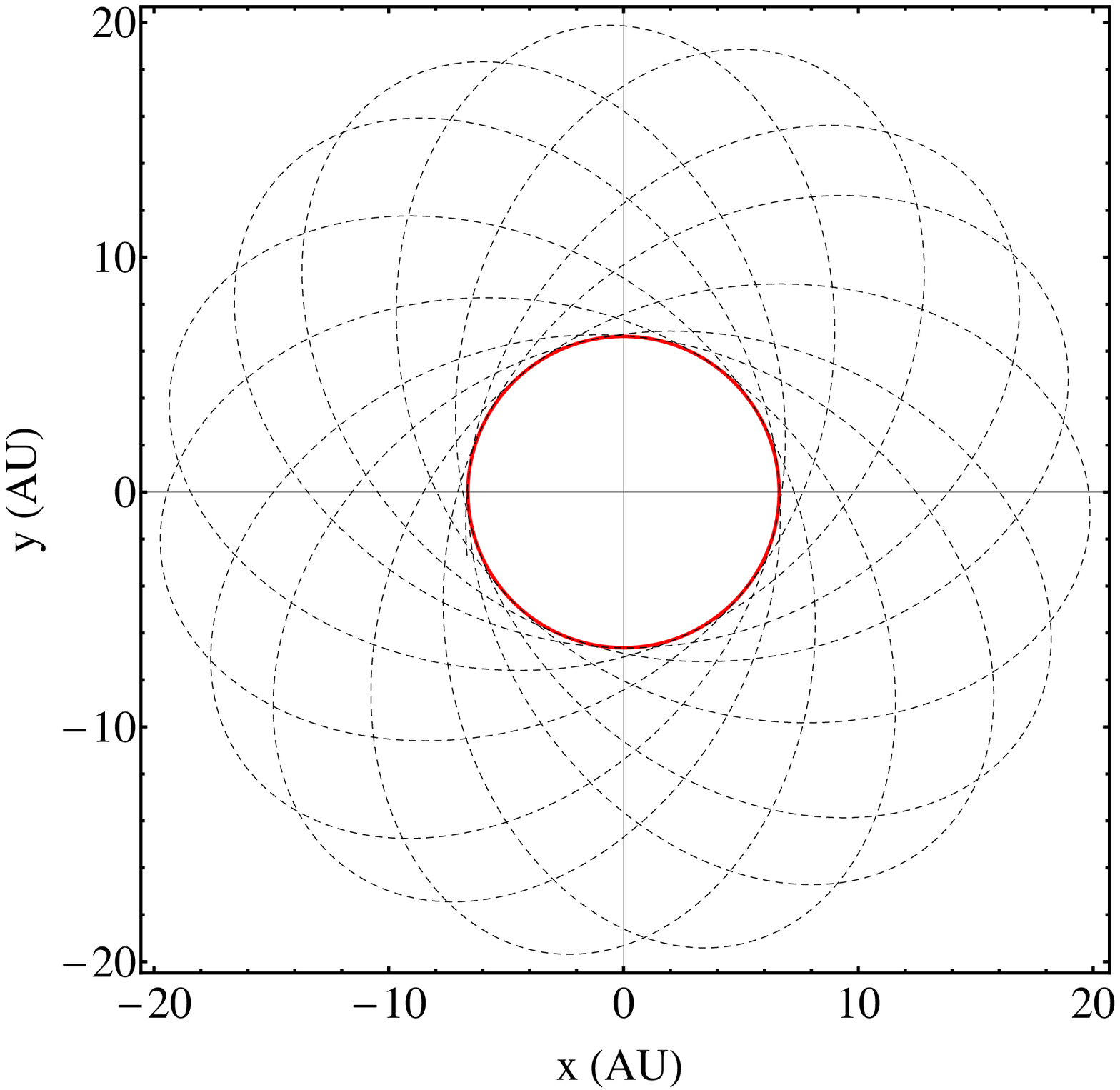} & \includegraphics[width=\columnwidth,clip]{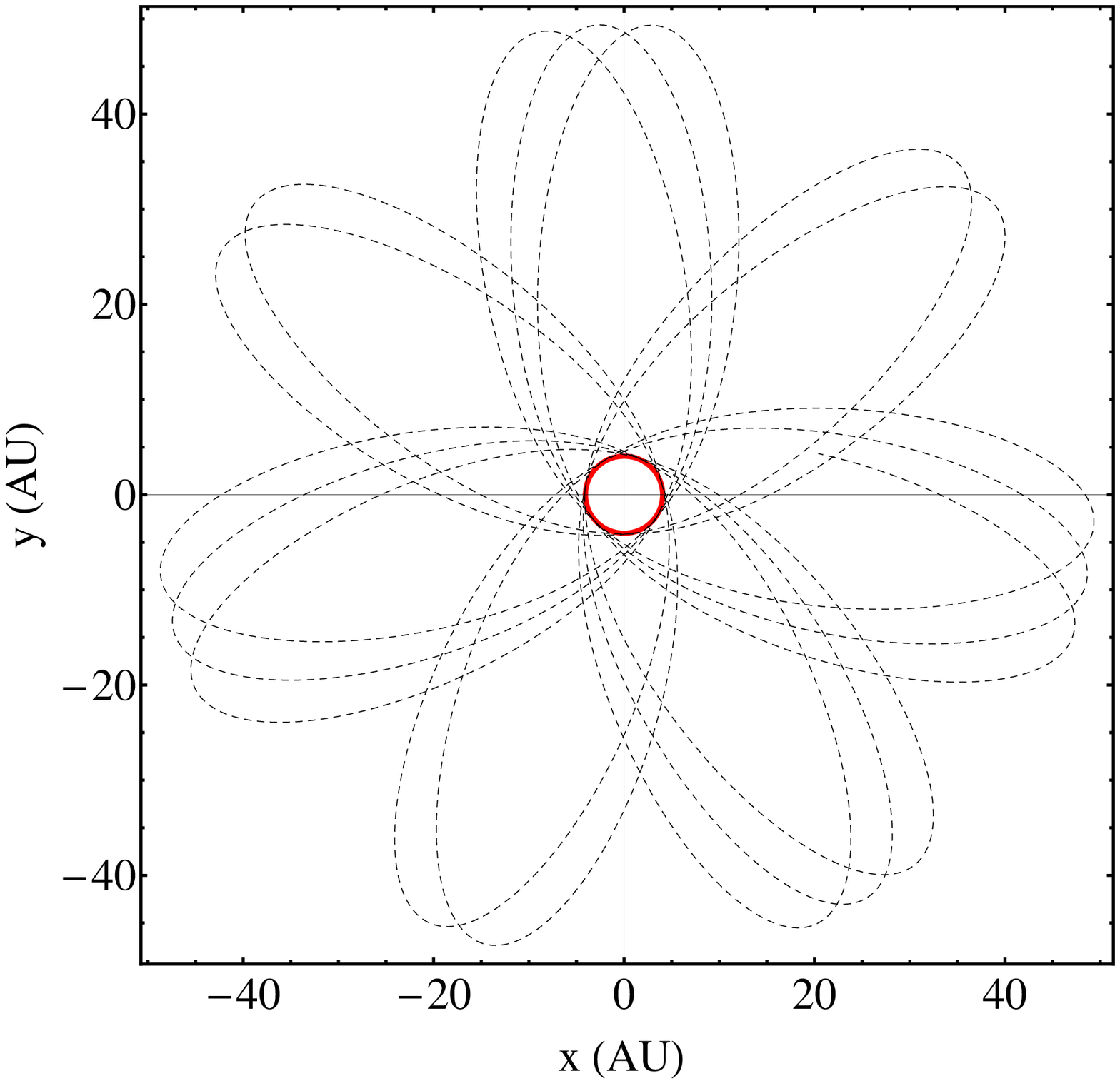}
\end{tabular}
\caption{(Color online) Geodesics near the galactic center. The solid red and the dashed black curves show the motion of test particles in the field of the SMBH and DM, respectively. Initial conditions are the same for both cases.  Left panel: $r(0)=80r_{BH}\approx6.63 AU, \dot r(0)=0, \phi(0)=0, \dot\phi(0)\approx2.38\times10^{-5}$ rad$/$s. Right panel: $r(0)=50r_{BH}\approx4.15 AU, \dot r(0)=0, \phi(0)=0, \dot\phi(0)\approx4.83\times10^{-5}$ rad$/$s.}\label{fig:y_vs_x_80RBH}
\end{figure*}

Making use of the metric functions and solving the geodesics equations numerically we compared the motion of test particles in the gravitational field of the SMBH with the corresponding geodesics in the field of the DM clump. Plots of the motion of test particles with the same initial energy and angular momentum are shown in Figs.~\ref{fig:y_vs_x_1000RBH} and \ref{fig:y_vs_x_80RBH}. The differences between the DM clump and the SMBH appear below 100 AU.
The initial conditions have been chosen in a such way to obtain circular geodesics in the field of the SMBH that lead to non circular orbits in the field of the DM cloud.

%%%%%%%%

The nearest observed star to Sgr-A$^*$ has its closest approach to the galactic center at roughly 120AU and therefore its motion cannot be used to distinguish between the two sources.
The discovery of a compact object, such as a pulsar, in the vicinity of Sgr-A$^*$ would allow to constrain the validity of the model without the need of optical observations of light coming from the accretion disk of the SMBH candidate. However at present, despite the theoretical prediction of the existence of such objects \citep{pulsars}, none has been observed.

%%%%%%%%%%%%%%%%%%%%%%%%%%%%%%%%%%%%%%%%%%%%%%%%%%%%%%%%%%%%%%%%%%%%%%%%%%%%%%%%%%%%%%%%%%%%%%%%%%%%%%%%%%%%%%%%%%%%%%%%%%%%%%%%%%%%%%%%%%%%%%%%%%%%%%%%%%%%%%%%%%%%%%%%%%%%%%%%%%%%%%%%
\section{Discussion and Conclusion}\label{sec:5}

We considered a toy model for a relativistic interior solution which can be used to model a low density distribution of matter in the inner part of the MWG. We considered the possibility that the proposed SMBH located in Sgr-A$^*$ may in fact be such a DM clump.
The density profile chosen is analogous to the exponential sphere that was introduced in
Ref.~\citep{sofue2013} to model DM in the galactic bulge.
The free parameters in the model were inferred by fitting the observed RC near the galactic center with the theoretical formula for the velocity obtained from the density profile.
The parameters give for the DM clump the same total mass as for the SMBH,
irrespective of the value of 'reasonable' constraints imposed on the size of the central object.

We studied the motion of test particles in the gravitational field
of both SMBH and DM core. As expected, a significant discrepancy in the motion appears only below 100 AU and it increases approaching the center. This means that current observations of the motion of stars, such as S2, that exist in the vicinity of Sgr-A$^*$ (see for example \citep{abuter2018}) can not distinguish our proposed DM clump from the SMBH. The discovery of a compact object, such as a pulsar, in the vicinity of Sgr-A$^*$ would allow to put better constraints on the models. However, as of now, no such pulsar has been observed, despite the theoretical prediction that there should exist hundreds of such objects within one parsec from Sgr-A$^*$
\citep{pulsars}.

Of course any realistic matter distribution would be rotating and possibly non isotropic, which would also cause departures from the behaviour described in the toy model. However, other measurements could in principle help to distinguish a black hole from another massive source.
Most notably, future observations of the galactic center via VLBI should be able to distinguish between different proposed scenarios due to the different structure of photon orbits in the vicinity of the expected location of the SMBH horizon.
In this context, the Black Hole Cam project \citep{BHC} aims at imaging the shadow of the compact objects located in Sgr-A$^*$ and at the center of the galaxy M87. It is important to study the properties of the shadow of alternatives to black holes in order to fully validate the conclusions that will be drawn from the expected observations. For example, the simulation of the shadow of a boson star at the galactic center was obtained in Refs.~\citep{boson-shadow1} and \citep{boson-shadow2} while that of a gravastar was obtained in Ref.~\citep{grava-shadow} and in both instances the exotic sources can be distinguished from a SMBH via observations obtainable with current technology.
As of now, we can not rule out the possibility that SMBH candidates are in fact exotic compact objects such as boson stars or gravastars. In a similar manner, we can not exclude the possibility that they may be simpler objects, for example, described by a matter profiles such as the one proposed here.

%%%%%%%%%%%%%%%%%%%%%%%%%%%%%%%%%%%%%%%%%%%%%%%%%%%%%%%%%%%%%%%%%%%%%%%%%%%%%%%%%%%%%%%%%%%%%%%%%%%%%%%%%%%%%%%%%%%%%%%%%%%%%%%%%%%%%%%%%%%%%%%%%%

\section*{Acknowledgement}
\medskip
\noindent
The authors thank the anonymous referee  for the useful and constructive comments, which helped to improve the presentation of the paper. The work was supported in part by Nazarbayev University Faculty Development Competitive Research Grants: 'Quantum gravity from outer space and the search for new extreme astrophysical phenomena', Grant No. 090118FD5348 and by the MES of the RK, Program 'Center of Excellence for Fundamental and Applied Physics' IRN: BR05236454, and by the MES Program IRN: BR05236494.

\bibliographystyle{mnras}
%\bibliography{biblio}

\end{document}